\documentclass[a4paper]{article}

\usepackage{graphicx}
\usepackage{amssymb}
\usepackage{amsmath}
\usepackage{natbib}     
                       
=10000

\title{Four-dimensional variational assimilation in the unstable subspace (4DVar-AUS) and the optimal subspace dimension}
\author{{Anna Trevisan}$^1$,  Massimo D'Isidoro$^1$ and {Olivier Talagrand}$^2$\\
$^1$ ISAC-CNR,  Bologna, Italy \\
$^2$ Laboratoire de M\'et\'eorologie Dynamique, ENS, Paris, France.
\\
\\ \LARGE{Submitted to Q. J. Roy. Met. Soc.}}

\begin{document}
\maketitle

\abstract

A key a priori information used in 4DVar is the knowledge of the system's
evolution equations.  In this paper we propose a method for taking full advantage of the
knowledge of the system's dynamical instabilities in order to improve the quality of the analysis. We
present an algorithm, four-dimensional variational assimilation in the unstable
subspace (4DVar-AUS), that consists in confining in this subspace the increment
of the control variable.  The existence of an optimal subspace dimension for
this confinement is hypothesized.  Theoretical arguments in favor of the
present approach  are supported by numerical experiments in a simple perfect
non-linear model scenario. It is found that the RMS analysis error is a
function of the dimension $N$ of the subspace where the analysis is confined
and is minimum for $N$ approximately equal to the dimension of the unstable and neutral
manifold.  For all assimilation windows, from 1 to 5 days, 4DVar-AUS performs
better than standard 4DVar. In the presence of observational noise, the 4DVar
solution, while being closer to the observations, if farther away from the
truth. The implementation of 4DVar-AUS does not require  the adjoint
integration.

\section{Introduction}
Accuracy in the definition of the initial condition is an important factor for
the performance of numerical weather and ocean prediction. The classical
problem of estimating the state of a dynamical system from noisy and incomplete
observations is known in meteorology and oceanography as {\em data assimilation}
\citep{daley-1991,kalnay-2003}. The goal of data
assimilation in the initialization process is to provide the best possible
estimate of the present state of the system using the available, partial and
noisy, observations and the approximate equations governing the system's
evolution.  The estimate, referred to as the {\em analysis}, is obtained by optimally
combining the information coming from a model forecast (background) and the
observations \citep{talagrand-jmsj-1997}.

The non-linear stability properties of the system do not only determine the
predictability horizon of the initial value problem but also profoundly
influence the assimilation process, affecting directly its quality and that
of the subsequent forecast \citep[see e.g.][and references therein]{carrassi_etal-chaos-2008}. All assimilation methods, more or less implicitly, exert some control on the flow-dependent instabilities by means of the observational information.
The Assimilation in the Unstable Subspace  \citep[AUS, ][]{trevisan_etal-jas-2004}
 explicitly estimates the flow-dependent instabilities and makes use of the 
unstable subspace as additional dynamical information.
The 4-dimensional extension of AUS is the main scope of the present paper.
In \citet{trevisan_etal-jas-2004} and in other applications of the AUS
assimilation, only a few unstable directions were tracked, whereas in the
present study we make use of the entire unstable and neutral subspace, the subspace 
spanned by the Lyapunov vectors with positive and null exponents.

Assimilation methods can be classified in
two categories: sequential and variational, the most notable in the two classes
being Kalman Filters and 4DVar respectively \citep[][and references
therein]{ghil_etal-ag-1991,kalnay-2003}. The Kalman Filter was originally
developed for linear systems but a straightforward way of extending the linear
results to the nonlinear case is given by the Extended Kalman Filter (EKF)
\citep{jazwinski-1970,miller_etal-jas-1994}.

Efficient minimization algorithms associated with adjoint techniques \citep{talagrand_etal_qjrms-1987} 
facilitate the implementation of 4DVar, an established and powerful assimilation method for
meteorology and oceanography.  
In many realistic circumstances, reduced-rank
approximations or a Monte Carlo approach, the latter referred to as Ensemble
Kalman Filter (EnKF) \citep{evensen-jgr-1994},  have been adopted to circumvent
the prohibitive cost of the full Extended Kalman Filter. The reader is referred
to \citet{kalnay-2003}  and \citet{tsuyuki_etal-jmsj-2007}  for a review on the
state of the art of data assimilation in meteorology; see also
\citet{lorenc-qjrms-2003}, \citet{kalnay_etal-tellus-2007} and
\citet{gustafsson-tellus-2007} for a discussion on the relative merits of 4DVar
and EnKF.

The system's unstable subspace and its role in the assimilation process
are central to our discussion. Hence, we briefly comment on how the flow
dependent instabilities are dealt with by Kalman type filters and
4DVar.  In the Kalman filter, the  propagation of the
flow dependent instabilities is obtained  by explicitly evolving the analysis
error covariance from the previous analysis step.  
%In its basic formulation, the EKF analysis error covariance ${\bf P}^a$ estimate remains confined within the same subspace of the forecast error covariance ${\bf P}^f$. 
In Ensemble Kalman
filters the subspace dimension of forecast error is at most equal to N-1,
if  N is the number of ensemble members: the rank deficiency of the background error information is partly
alleviated by covariance localization if N is too small
\citep{hamill_etal-mwr-2001}. A related aspect is the filter divergence that
appears particularly critical in relation to sampling error \citep{whitaker_etal-mwr-2002}.  

4DVar generates a model trajectory that best fits the
observations available within a given assimilation window.
Within the assimilation window, the flow dependent instabilities are naturally described by the forward
integration of the model and backward integration of the adjoint that model the
error evolution.
In addition, at the start of each assimilation window, an a priori estimate of the
background error covariance is needed \citep{bannister-qjrms-2008}.

For long assimilation windows, 4DVar analysis errors are known to project on the
unstable subspace of the system \citep{pires_etal-tellus-1996}. Errors in the
stable directions that would be damped in the long range, for short
assimilation windows are non-negligible in the analysis and affect the next
assimilation cycle, causing short term enhanced error growth
\citep{swanson_etal-jas-2000}.
It therefore seems appropriate to avoid introducing such type of error: this
can be achieved by confining the increment of the 4DVar control variable in the
unstable and neutral subspace of the system. 
In this way we avoid reintroducing observational error in the stable directions
at each assimilation step: we anticipate that this is beneficial only if
observations are not perfect.
In this paper we present an algorithm (4DVar-AUS) that minimizes the 4DVar cost
function under the above constraint. The dynamical information on the growth of
errors in the unstable and neutral directions, the Lyapunov vectors with positive and null exponents, 
is explicitly estimated and, as explained in Section \ref{2.2.2}, the adjoint integration is not needed. 

The idea of confining the analysis increment in the unstable subspace is not
new. The sequential algorithm, referred to as AUS, has been introduced by \citet{trevisan_etal-jas-2004}. Its application to different models
and observation configurations has shown that, even in the context of
high-dimensional systems, an efficient error control and accurate state
estimate can be obtained even by monitoring only a reduced number of unstable
directions \citep{uboldi_etal-npg-2006,carrassi_etal-tellus-2007,carrassi_etal-npg-2008}.
The basic elements of the AUS scheme that differentiate it from other  ensemble type Kalman filters are the explicit monitoring of the unstable directions of the
system and the confinement of the analysis increment in the subspace that they span. 
%In a perfect model, a number of perfect observations equal to the number of Lyapunov positive exponents is sufficient to determine the state of the system 
%\citep[see e.g.][]{trevisan_etal-jas-2004,carrassi_etal-chaos-2008}. 
A localization of the unstable structures, and consequently a localization of the 
error covariances (a feature common to other EnKF type methods) is necessary if
the dimension of the subspace for the AUS assimilation is too small to
describe the background error.
The present extension of AUS to the four-dimensional case has the advantage of
using the time distributed observations to track the instabilities that develop
along the flow.

One of the main goals of the present study is to address the following
question: is there an optimal subspace dimension for the  assimilation and is
this related to the dimension of the  unstable subspace of the system? In order
to address this question, 4DVar-AUS is formulated in a perfect model setting
and using a subspace of variable dimension.

Theoretical arguments will be presented to indicate that the subspace dimension
should at least be equal to the unstable manifold dimension.

%In the present formulation of 4DVar-AUS, the information on the flow dependent
%instabilities is propagated from one cycle to the next; the most unstable
%directions of the best fitting trajectory are used, in the next cycle, to
%confine the analysis increment in their subspace (4DVar-AUS). In this way,
%errors in the stable directions do not influence the analysis even for short
%assimilation windows. This is beneficial for different reasons: the analysis e
%sull'attrattore, elimina crescita transiente, e' ben condizionata,
%numericamente vantaggiosa non usa l'aggiunto. 

Results of the application of 4DVar-AUS to simple perfect non-linear systems
obtained by varying the  number of degrees of freedom in the Lorenz $40$-variable model \citep{lorenz-1996} 
will be presented. The relation between the optimal subspace dimension
and the number of positive exponents will be numerically investigated  and it
will be shown that the results confirm the theoretical arguments.

The paper is organized as follows: the formulation of 4DVar-AUS and theoretical
arguments on the subspace optimization are introduced in Section \ref{4dvar-aus};
results of the application to the \citet{lorenz-1996} model are presented in
Section \ref{application}, while conclusions are drawn in Section \ref{conclusions}.

\section{Formulation of 4DVar in the unstable subspace}
\label{4dvar-aus}
\subsection{4DVar}
Strong constraint 4DVar seeks the (nonlinear) best estimate of the initial
state ${\bf x}_0$ that minimizes the misfit with observations in a given time
interval (window) and possibly with a background state ${\bf x}_0^b$.  The
standard cost function for strong constraint 4DVar, in discrete form, can
be written as:

\begin{equation}
\label{cost_function}
\begin{split}
 J({\bf x}_0)&= ({\bf x}_0-{\bf x}^b_0)^T {\bf B}^{-1}({\bf x}_0-{\bf x}^b_0)+ \\
&+\sum_{i=0}^{n} ({\cal H}_i{\bf x}_i-{\bf y}_i^o)^T  {\bf R}^{-1} ({\cal H}_i{\bf
x}_i-{\bf y}_i^o) 
\end{split}
\end{equation}
 where ${\bf y}^o_i$ are the observations available at discrete
times $t_i=i{\Delta t}, i=0,...,n$ , within the assimilation window  
of length $\tau=t_n-t_0$; $\bf B$  and $\bf R$
represent the background  and observation error covariance matrices,  ${\cal H}$ 
the nonlinear observation operator, and the sequence of model states ${\bf x}_i$ 
is a solution of the nonlinear model equations:
\begin{equation}
{\bf x}_i= {\cal M}_{0\rightarrow i} ({\bf x}_0),
\label{model_traj}
\end{equation}

The control variable for the minimization is the model state  ${\bf x}_0$ at
the beginning of the assimilation window.

Given the tangent linear equations describing the evolution of
infinitesimal perturbations $\delta{\bf x}_i$ relative to an orbit of
Eq. (\ref{model_traj}):
\begin{equation}
\delta{\bf x}_i = {\bf M}_{0\rightarrow i}\delta{\bf x}_0,
\end{equation}

the gradient of $J$ with respect to  ${\bf x}_0$ can be written as:

\begin{equation}
\label{grad_J}
\begin{split}
 \frac{1}{2}\nabla_{{\bf x}_0} J&={\bf B}^{-1}({\bf x}_0-{\bf x}^b_0)+\\
&+\sum_{i=0}^{n} {\bf M}_{0\rightarrow i}^T {\bf H}_i^T {\bf R}^{-1} ({\cal H}_i{\bf x}_i-{\bf y}_i) 
\end{split}
\end{equation}

where ${\bf H}_i$ represents the linearized observation operator, and the superscript $T$ stands for transpose.

For a given nonlinear trajectory, the gradient can be estimated by use of the
adjoint method \citep{ledimet_etal-tellus-1986}.

The solution of the minimization problem is
obtained by forward integration of the model and backward integration of the
adjoint, with an iterative descent algorithm.

\subsection{4DVar-AUS}
\label{4dvar-AUS}

\subsubsection{Unstable subspace and computation of Lyapunov vectors}
\label{2.2.1}

Lyapunov vectors, defined for nonlinear systems, are the time dependent
physical structures associated with the Lyapunov exponents. There are basically
two definitions of Lyapunov vectors that span the same invariant
Oseledec subspaces. The first is that of an orthonormal set of vectors, the 
eigenvectors of the limit operator \citep{oseledec-1968}:
\begin{equation}
{\bf \Phi}_{\infty} (t)=\lim_{t_0 \rightarrow  -\infty} [ {\bf M} _{t_0 
\rightarrow t}{\bf M}_{t_0 \rightarrow t}^*] ^{\frac{1}{2 (t-t_0)}}
\end{equation}
where ${\bf M}^*$ is the adjoint operator and the initial state, ${\bf x}_0$ of
the nonlinear trajectory is on the attractor (for further reference in the
meteorological literature see \citet{lorenz-physica-1984} and
\citet{legras_etal-1996}).

The second is a set of non-orthogonal Lyapunov vectors, that are independent of the norm
and map into themselves with the tangent linear propagator
\citep{eckmann_etal-rmp-1985,brown_etal-pra-1991}.  These vectors have been shown to be the natural
generalization of eigenvectors and Floquet vectors to aperiodic flow
\citep{trevisan_etal-jas-1998}.

The following standard technique is commonly used for calculating the 
orthonormal set of Lyapunov vectors \citep{benettin_etal-meccanica-1980}: 
a set of $N$ initially random  tangent vectors are linearly  evolved and
orthonormalized every $\tau$ time units.  After a spin-up time these vectors
span the $N$-dimensional most unstable subspace of the system.  

An efficient method for recovering norm-independent non-orthogonal Lyapunov
vectors is given by \citet{wolfe_etal-tellus-2007}.  Either one of the two
above-mentioned methods can be used to identify the unstable, neutral, stable
subspaces, the span of Lyapunov vectors with positive, null, negative
exponents; the former, simpler technique will be adopted in the present
application.

In weather prediction, bred vectors \citep{toth_etal-bams-1993, toth_etal-mwr-1997} are usually 
computed instead of Lyapunov vectors. Bred vectors are the finite amplitude generalization
of Lyapunov vectors and are computed as differences between twin
nonlinear model integrations. The re-normalization amplitude and breeding time
are the parameters being tuned to select the instability scale. With an
infinitesimal re-normalization amplitude and periodic orthonormalization 
the bred vectors algorithm would produce the same results as the Lyapunov vectors algorithm
of \citet{benettin_etal-meccanica-1980}.

In previous applications of AUS \cite[see e.g.][and references therein]{carrassi_etal-npg-2008} to more realistic atmospheric and oceanic
models, the unstable vectors were computed with the breeding technique. In those
works, only a small number of bred vectors was used at each assimilation time.
Because of the low dimensionality of the subspace spanned by those bred vectors a
space localization was needed; furthermore, the breeding method was applied to
the assimilation system instead of the freely evolving system in order to
select those instabilities that survived the previous assimilation.

In the present application we use the entire unstable and neutral subspace of the freely
evolving system and we do not need any localization.

\subsubsection{The 4DVar-AUS algorithm}\label{2.2.2}
The approach consists in determining the increment 
$\delta{\bf x}_0$ which minimizes the cost function in the 
reduced dimension subspace spanned by the $N$ most unstable 
directions of the system corresponding to the leading $N$ Lyapunov exponents.

After a transient time, the numerical technique described in Section \ref{2.2.1}
\citep{benettin_etal-meccanica-1980} leads to the identification of an
orthonormal set of vectors spanning the $N$-dimensional most unstable subspace.
We apply this procedure to the solution of the assimilation cycle, starting
initially with $N$ arbitrarily chosen tangent vectors. The Gram-Schmidt
orthonormalization is applied at the end of each assimilation interval $\tau$.

Let ${\bf E}_0$ be the matrix whose columns are the orthonormal tangent vectors
spanning the $N$-dimensional most unstable subspace of the system at $t_0$. The
linear evolution within the assimilation window $[t_0,\tau]$ is given by:

\begin{equation}
\label{e0_evolution}
{\bf M}_{0\rightarrow i}{\bf E}_0={\bf E}_i{\bf\Lambda}_{i} , (i=0,...,n)
\end{equation}
where 
\begin{equation}
\begin{split}
{\bf \Lambda}_{i}&=diag \biggl[exp\int_{t_0}^{t_i}\lambda ^{(1)}(t) dt,
exp\int_{t_0}^{t_i}\lambda ^{(2)}(t) dt,...,\\
&,..., exp\int_{t_0}^{t_i}\lambda ^{(N)}(t)dt\biggr]
\end{split}
\end{equation}
and  $\lambda^{(j)}(t)$ is the $j^{th}$  local  exponent.

Consider the projection of the increment $\delta{\bf x}_0$  in the subspace
defined by ${\bf E}_0$. In general, given a norm defined by the symmetric
positive definite matrix ${\bf Q}$, the projection can be written as 
${\bf E}_0({\bf E}^T_0{\bf Q}{\bf E}_0)^{-1}{\bf E}_0^T{\bf Q}\delta{\bf x}_0$. For
simplicity, in the following we adopt the Euclidean norm and we recall that the
columns of ${\bf E}_0$ are orthonormal vectors.

Thus, let the increment $\delta{\bf x}_0$ be confined in the subspace ${\bf E}_0$ and its projection  $\widetilde{\delta{\bf x}}_0$  be given by:
\begin{equation}
\label{projection}
\widetilde{\delta{\bf x}}_0={\bf E}_0{\bf E}_0^T \delta {\bf x}_0
\end{equation}
The evolution of the projected increment is governed by:
\begin{equation}
\label{dxi}
 \widetilde{\delta {\bf x}}_i={\bf M}_{0\rightarrow i}{\bf E}_0{\bf E}_0^T \delta{\bf x}_0=
{\bf E}_i{\bf \Lambda}_{i}{\bf E}_0^T \delta {\bf x}_0 , (i=0,...,n)
\end{equation}
Variations of the cost function (\ref{cost_function}) due to variations of the control variable $\widetilde{\delta{\bf x}}_0$ can be written as:
\begin{equation}
\label{j_tilde}
\begin{split}
\widetilde{\delta {\bf J}}&=(\nabla_{x_0}J)^T\widetilde{\delta{\bf x}}_0\\
&=(\widetilde{\nabla_{x_0}J})^T{\bf \delta x}_0
\end{split}
\end{equation} 

where the tilde represents the projection into the subspace ${\bf E}_0$, \textit{i.e.} 
$\widetilde{\nabla_{x_0}J}={\bf E}_0{\bf E}_0^T\nabla_{x_0}J$.

Using (\ref{grad_J}) and (\ref{e0_evolution}), the cost function gradient in
the reduced subspace becomes: 
\begin{equation}
\label{gradient}
\begin{split}
\frac{1}{2}\widetilde{\nabla_{{\bf x}_0}J}&={\bf E}_0\biggl[{\bf E}_0^T{\bf B}^{-1}({\bf x}_0-{\bf x}^b_0)+\\
&+\sum_{i=0}^n{\bf \Lambda}_i{\bf E}_i^T{\bf H}_i^T{\bf R}^{-1}(\mathcal{H}_i{\bf x}_i-{\bf y}_i)\biggr]
\end{split}
\end{equation}

An assimilation cycle is constructed by initializing the next assimilation window
with the state and its associated unstable subspace estimates at the end of the
previous window. 

If $k$ indicates the index of the assimilation cycle and recalling that $\tau=t_n-t_0$:

${{\bf x}^b_0}(k+1)={\bf x}_{\tau}(k)$, where
\begin{equation}
\label{xtau}
{\bf x}_{\tau}(k)= {\cal M}_{0\rightarrow n}(k) ({\bf x}_0(k))
\end{equation}
and ${\bf E}_{\tau}(k)={\bf E}_{0}(k+1){\bf T}$, where
\begin{equation}
\label{etau}
{\bf M}_{0\rightarrow n}(k){\bf E}_0(k)={\bf E}_{\tau}(k){\bf\Lambda}_{\tau}(k)
\end{equation}
and ${\bf T}$ is the upper triangular orthonormalization matrix.

In summary, the assimilation cycle is performed through the following steps:
\begin{enumerate}
\item  A descent algorithm is used to find the cost function minimum by:
\begin{enumerate}
\item forward integration of the nonlinear model using (\ref{model_traj}) to compute
${\bf x}_i$, starting from ${\bf x}_0^b$ at first iteration step; 
\item forward integration of the perturbations ${\bf E}_0$ using (\ref{e0_evolution}) to compute 
${\bf E}_i$ and ${\bf \Lambda}_i$;
\item estimate of $\widetilde{\nabla_{{\bf x}_0} J}$ from (\ref{gradient}) and of $J$ from (\ref{cost_function});
\end{enumerate}
\item The nonlinear model is integrated starting from  the minimizing solution ${\bf x}_0(k)$ 
to produce the analysis, ${\bf x}_{\tau}(k)$ (\ref{xtau}).
\item The perturbations ${\bf E}_0(k)$ are evolved along the minimizing trajectory to 
produce ${\bf E}_{\tau}(k)$ (\ref{etau}); 
\item The columns of ${\bf E}_{\tau}(k)$ are orthonormalized and stored in ${\bf E}_0(k+1)$ 
to be used in the next assimilation cycle; 
\item ${\bf x}_0^b(k+1)$ is set equal to ${\bf x_{\tau}}(k)$.
\end{enumerate}

Notice that no use of the adjoint integration is made.

In the 4DVar-AUS assimilation, the analysis increment is confined in the
$N$-dimensional most unstable subspace of the previous analysis solution, with $N$
approximately equal to the number of positive and null Lyapunov exponents.

Theoretical arguments given in Section \ref{teoria}, confirmed by numerical results presented 
in Section \ref{application}, will show that during  a 4DVar-AUS assimilation cycle,
errors in the stable directions are damped and errors in the analysis solution
are confined within the unstable and neutral manifold of the system.  This subspace is
locally parallel to the attractor \citep{eckmann_etal-rmp-1985}, so that one can find
a state belonging to a nearby trajectory by moving along the tangent unstable directions.

\subsection{Full space and reduced order covariance matrix of the assimilation error}
\label{teoria}
The effect of the confinement on the expected assimilation error covariance is now examined.

\citet{pires_etal-tellus-1996}  investigated the behavior of the observational
term of the cost function in chaotic systems, making the tangent linear
hypothesis and observing the whole state.   They showed that, using the
assumption that the observation error is uncorrelated in time and isotropic,
with variance $\sigma_o^2$, the covariance matrix ${\bf C}_0 = <{\bf \eta}^a_0{\bf \eta}^{aT}_0>$ 
of the assimilation error ${\bf \eta}^a_0$ at
$t=t_0$, $< >$ being the expectation operator, can be written as:

\begin{equation}
\label{c0_pires}
{\bf C}_0= \sigma_o^2 \left(\sum_{i=0}^{n}{\bf M}_{0\rightarrow i}^T {\bf M}_{0\rightarrow i}\right)^{-1} 
\end{equation}

By confinement in the subspace defined by the $N$ column vectors in ${\bf E}_0$ and using (\ref{e0_evolution}) one easily obtains the following expression for the covariance of the assimilation error:

\begin{equation}
{\bf C}_0 = \sigma_o^2{\bf E}_0(\sum_{i=0}^{n} {\bf \Lambda}_i {\bf E}_i^T{\bf E}_i {\bf \Lambda}_i)^{-1}{\bf E}_0^T
\label{c0_nostra}
\end{equation}
To this point, no hypothesis has been made on the choice of ${\bf E}_0$ in (\ref{c0_nostra}).  If
the number $N$ of vectors of ${\bf E}_0$ is equal to the total number, $I$, of 
degrees of freedom of the system, (\ref{c0_nostra}) represents the covariance matrix in the full 
space.

Now let the $N$ column vectors of ${\bf E}_0$ be the Lyapunov vectors corresponding to
the $N$ largest Lyapunov exponents. Assume that the Lyapunov vectors are orthogonal
at $t_0$ (or have been orthonormalized) and assume, for the sake of simplicity,
that they remain orthogonal within the time window, then:
\begin{equation}
\label{c0}
\begin{split}
&{\bf C}_0=  \sigma_o^2 {\bf E}_0 {\bf D}_0 {\bf E}_0^T\\
&= \sigma_o^2{\bf E}_0 diag\biggl\{\bigl[\sum_{i=0}^{n}( {\bf \Lambda}_i^{(1)})^2\bigr]^{-1}, \bigl[\sum_{i=0}^{n}( {\bf \Lambda}_i^{(2)})^2\bigr]^{-1},...,\\
&,...,\bigl[\sum_{i=0}^{n}( {\bf \Lambda}_i^{(N)})^2\bigr]^{-1}\biggr\}{\bf E}_0^T
\end{split}
\end{equation}
where  ${\Lambda}_i^{(j)} =exp\int_{t_0}^{t_i}\lambda^{(j)}(t) dt$.

At time $\tau=t_n$, the analysis error covariance $C_{\tau} = <{\bf \eta}^a_{\tau}{\bf \eta}^{aT}_{\tau}>$ under the tangent linear assumption, is: 
\begin{equation}
\label{ctau}
\begin{split}
&{\bf C}_{\tau} =  \sigma_o^2 {\bf E}_{\tau} {\bf D}_{\tau} {\bf E}_{\tau}^T\\
& = \sigma_o^2{\bf E}_{\tau} diag \biggl\{ \bigl[\sum_{i=0}^{n} ({\bf \Lambda}_i^{(1)})^{-2}\bigr]^{-1},
\bigl[\sum_{i=0}^{n} ({\bf \Lambda}_i^{(2)})^{-2}\bigr]^{-1},...,\\
&,...,\bigl[\sum_{i=0}^{n} ({\bf \Lambda}_i^{(N)})^{-2}\bigr]^{-1}\bigg\}{\bf E}_{\tau}^T
\end{split}
\end{equation}
 where ${\bf E}_{\tau}={\bf E}(t_n)$.

In the expressions for ${\bf C}_0$ and ${\bf C}_{\tau}$, the role of the amplifying and decaying modes
is interchanged.  The generic term $(D_0)_{j,j}$ of the
diagonal matrix 
${\bf D}_0$, representing the analysis error covariance associated with 
the $j^{th}$ Lyapunov direction (the $j^{th}$ column of ${\bf E}(t_0)$, ${\bf e}_j(t_0)$)
at the beginning of the assimilation window, $t=t_0$ is:
\begin{equation}
\label{expt0}
\begin{split}
 (D_0)_{j,j}= \sigma_o^2\bigl\{ 1+[exp(\lambda^{(j)} \Delta t)]^2 &+[exp(\lambda^{(j)} 2\Delta t)]^2 +\\
&+...+[exp(\lambda^{(j)} n\Delta t)]^2 \bigr\}^{-1}\\
&= \sigma_o^2\frac{1-e^{2\lambda^{(j)} \Delta t}}{1-e^{2(n+1)\lambda^{(j)} \Delta t}}
\end{split}
\end{equation}
 where $\lambda ^{(j)}$ is the  $j^{th}$ local Lyapunov exponent assumed, to simplify notation, to be constant within the assimilation window and $\Delta t$ is the time interval between observations.
 A similar expression holds at the end of the assimilation window, $t=\tau$:
\begin{equation}
\label{exptau}
\begin{split}
  (D_{\tau})_{j,j}= \sigma_o^2\bigl\{ 1+[exp(-\lambda^{(j)} \Delta t)]^2& +[exp(-\lambda^{(j)} 2\Delta t)]^2 +\\
&+...+[exp(-\lambda^{(j)} n\Delta t)]^2 \bigr\}^{-1}\\
&=\sigma_o^2 \frac{1-e^{-2\lambda^{(j)} \Delta t}}{1-e^{-2(n+1)\lambda^{(j)} \Delta t}}
\end{split}
\end{equation}
where, now, large $n$ refers to earlier times.

In agreement with \citet{pires_etal-tellus-1996}, the influence of observational error on the stable (unstable) directions is damped as time increases (decreases) within the assimilation interval.
At $t=\tau$ ($t=t_0$), the largest error is along the most unstable (stable) directions; 
the older (more recent) observations, corresponding to increasing $n$, give smaller and smaller contributions.
For  sufficiently long assimilation windows the error along the stable (unstable) directions is damped.

\subsection{On the optimal subspace dimension}
Now, we focus on the effect on the analysis error covariance of the
assimilation in the reduced, unstable and neutral subspace.
In 4DVar-AUS, the analysis increment is confined in 
the subspace of the $N$  most unstable directions: let $N$ be
equal to $N^+ +N^0$, where $N^+$ and $N^0$ are the number of positive and
null Lyapunov exponents.
The influence of errors in stable directions,
${\bf e}_{N^+ +N^0 +1},{\bf e}_{N^+ +N^0 +2},..{\bf e}_{I}$ on the analysis error
is eliminated by the confinement.
Because we do not make corrections in the stable directions we avoid
introducing, at every  assimilation step, errors in the assimilation solution
that project on the stable subspace. In this way, errors in the stable
directions can naturally decay along the cycle.  In standard 4DVar, instead, errors in
the observations produce errors in the assimilation solution that project on
the stable directions at each assimilation step.

Results presented in the next section confirm these arguments, showing that the
confinement in the unstable subspace is indeed  beneficial to the performance
of the assimilation.  At final time of sufficiently long assimilation windows
analysis errors in the stable directions would be damped also in standard
4DVar; however, this cannot be achieved in practice because nonlinearities set
a limit to the window extension that can be used.

\citet{swanson_etal-jas-2000} investigated the effect of various observational
errors on the 4DVar analysis and pointed out that errors in the stable
directions can cause short term enhanced growth with adverse effect on the
forecast. Their results are in agreement with the present findings.

The effect of the confinement on the efficiency of the 
numerical algorithm is discussed in Appendix \ref{appendice}.  

\section{Application to the Lorenz (1996) model}
\label{application}
\subsection{Model and experimental setting}
The results presented in this section are based on the low-order chaotic  model of \citet{lorenz-1996} and \citet{lorenz_etal-jas-1998}. The model is a simple analogue of mid-latitude atmospheric dynamics and its variables represent the values of a meteorological quantity at $I$ equally spaced geographic sites on a periodic latitudinal domain.

\noindent
The governing equations are:
\begin{equation}
\frac{d}{dt}x_j=(x_{j+1}-x_{j-2})x_{j-1}-x_j +F
\end{equation}

with $j=1,...,I$.

Following \citet{lorenz_etal-jas-1998} we set the external forcing $F=8$, a
value giving rise to chaotic behavior. In this paper we consider three model 
configurations with different numbers of degrees of freedom $I$. For $I=40, 60, 80$ the
three systems have $13, 19, 26$ positive Lyapunov exponents, respectively. The
doubling time associated to the leading Lyapunov exponent $\lambda^{(1)}$ is,
in all three systems, approximately equal to $2$ days if the system time unit
corresponds to $5$ days.

\begin{figure*}
\includegraphics[scale=0.435]{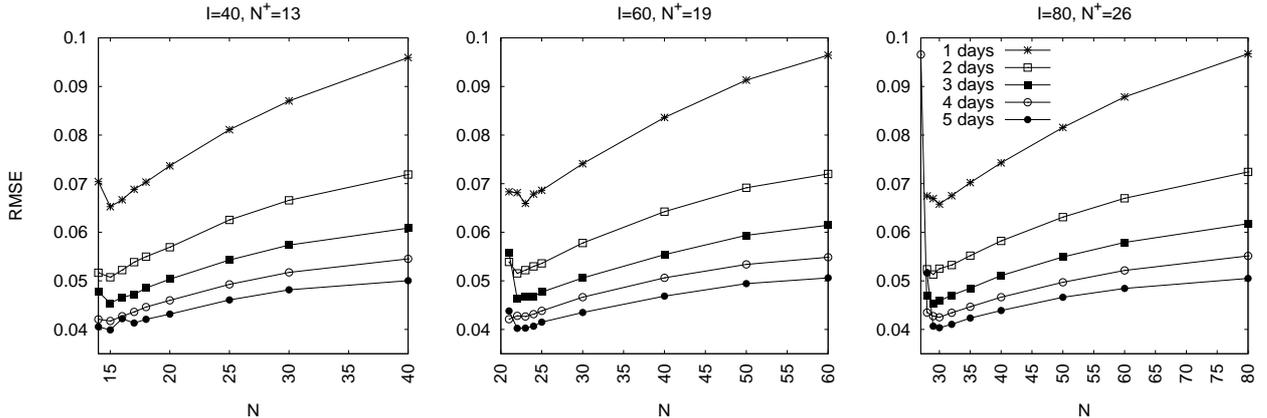} 
\caption{Time average RMS analysis error at $t=\tau$ as a function of the subspace dimension $N$ for three model configurations: $I$=$40$, $60$, $80$.  Different curves in the same panel refer to different assimilation windows from 1 to 5 days. The observation error standard deviation is $\sigma_o=0.2$.}
\label{fig1}
\end{figure*}

Observing system simulation experiments are performed in a perfect model
environment: a trajectory on the attractor of the system is assumed to
represent the {\em truth}.  Observations are created by adding to the {\em
true} state Gaussian distributed random errors with variance $\sigma_o^2$.

The observational network is the same as in \citet{fertig_etal-tellus-2007}.
An observation is placed in one out of four grid points at each observation
time. The frequency of observation is $1.5$ hours, and the observed grid points
are rotated so that, in a six hours interval, all grid points are observed
once.

An analysis cycle is set up with contiguous assimilation windows so that  the
initial time $t_0$  in one window corresponds to the last time $t_n =\tau$ of
the previous one, as described in Section \ref{2.2.2}. Experiments are
performed using assimilation windows $\tau=1,...,5$ corresponding to  $1,...,5$
days. In each experiment the analysis cycle consists of $5000$ consecutive
windows. A {\em Conjugate Gradient} algorithm \citep[][Chap. 10]{num_recipes}
is used for the minimization of the cost function at each step of the algorithm
of both 4DVar and 4DVar-AUS.

The time mean, over an assimilation cycle of $5000$  windows, of the RMS
analysis error obtained with standard 4DVar is compared with that obtained by
means of the 4DVar-AUS algorithm. The latter is applied using  a variable
number $N$ of directions in order to find the optimal subspace dimension for
the confinement.

\subsection{Results}
Following the theoretical approach of \citet{pires_etal-tellus-1996}, a first set
of experiments is performed without background term in the cost function. The
4DVar-AUS algorithm described in Section \ref{2.2.2} requires the definition of the background 
(i. e. the estimate produced by the previous assimilation cycle) and of the subspace ${\bf E}_0$
representing instabilities at the end of the trajectory in the previous
window. Therefore, even in the absence of an explicit background term in the cost
function, the solution relative to one window is dependent on the solution of
the previous one. 
For the first set of experiments an assimilation cycle is set up by 
successive minimizations of the observational part of the cost function only.
We recall that the initial guess of the control variable at $t=t_0$ is equal to
the analysis at $t=\tau$ of the previous assimilation and, in the 4DVar-AUS
algorithm, the $N$ column vectors of ${\bf E}_0$ are obtained by
orthonormalizing the vectors ${\bf E}_{\tau}$. 

The results of 4DVar-AUS are compared with standard 4DVar.
The RMS analysis error is computed at the
end of each assimilation window and averaged over $5000$ successive
assimilation windows. Experiments are performed with the three systems with
$I=40, 60$ and $80$ degrees of freedom; the number of null Lyapunov exponents of these
systems is $N^0=1$ and the number of positive exponents is
$N^+=13,19,26$, respectively.
\begin{figure*}
\includegraphics[scale=0.435]{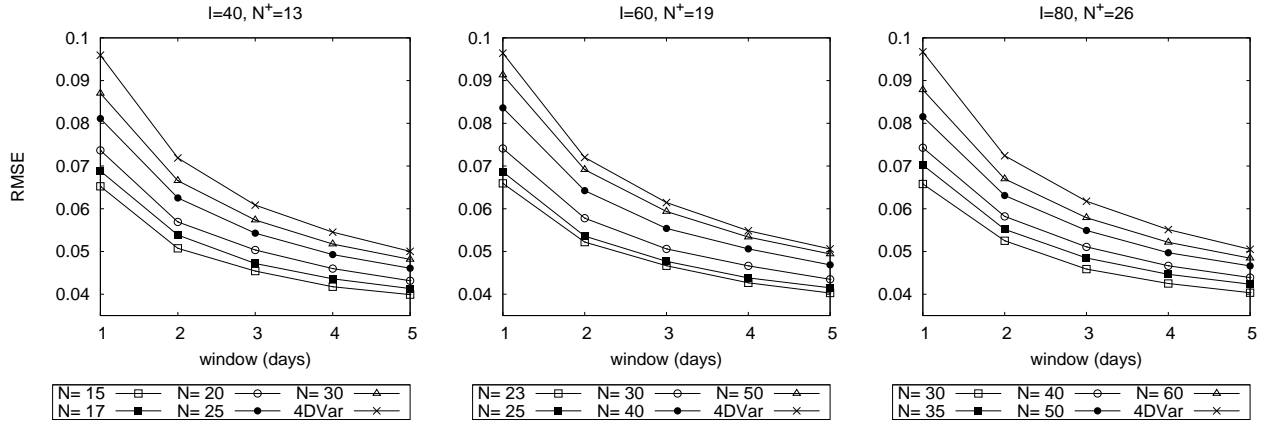} 
\caption{Time average RMS analysis error at $t=\tau$ as a function of the
length of the assimilation window for three model configurations: $I$=$40$,
$60$, $80$. Different curves in the same panel refer to a different subspace
dimension $N$ of 4DVar-AUS  and to standard 4DVar. $\sigma_o=0.2$.}
\label{fig2}
\end{figure*}
\subsubsection{Error dependence on the subspace dimension}
Figure \ref{fig1} shows the mean RMS
error as a function of the dimension $N$ of the subspace ${\bf E}_0$. Each 
panel refers to a different model configuration, $I=40, 60, 80$. 
When $N=I$ the error is that of standard 4DVar (one can either set $N=I$ in the
4DVar-AUS scheme or use the standard 4DVar algorithm: the results are the same
within numerical accuracy). The value of $\sigma_o$ is set to $0.2$ (the
'climatological' standard deviation for the system is about $5.1$).

When $N$ is smaller than the number of positive Lyapunov exponents, $N^+$, 
the 4DVar-AUS algorithm does not converge or gives very poor results. When
$N$ is increased above this threshold, the error abruptly decreases and then
gradually increases again up to the value obtained with standard 4DVar ($N=I$).
Recalling that the number of positive global Lyapunov exponents for $I=40, 60$
and $80$ is $13, 19$ and $26$ respectively, the error minimum is obtained in
all three model configurations for a value of $N$ approximately equal to
$N^+ +N^0$.  Because the value of local Lyapunov exponents fluctuates
around the respective global value, even moderately decaying directions can be
locally expanding and a number $N$ a few units larger than $N^+ +N^0$ is needed.

The most important result is that the minimum value of the error is obtained
for an optimal subspace dimension which is very close to the number of positive
and null Lyapunov exponents of the three ($40, 60$ and $80$-variable) systems.
 
Notice the internal consistency of the results of Fig. \ref{fig1}: 
the value of the average RMS error is virtually the same in the three model
configurations. In fact, dynamically, the three models are equivalent and have
the same instabilities, but a different number  $N^+$ of unstable directions
are present in proportion to the extension of the spatial domain. Because the
observational configuration is the same, with a number of observations
proportional to the domain size, and using the value of $N$ appropriate for
each system (given the respective value of $N^+$), we obtain the same
accuracy of the analysis solution.

Figure \ref{fig2}  displays the experimental data as a function of $\tau$, to
illustrate the improvement obtained with the 4DVar-AUS scheme for different
assimilation windows. Because the stable directions have a negative impact on
the quality of the assimilation, the error appears to decrease by successively
discarding a larger number of (the most stable) directions. We argued that
errors in the stable directions do not affect the analysis for long enough
(relative to the decay rate) assimilation windows. In agreement
with this conjecture, Fig. \ref{fig2} shows that the improvement obtained by
using the optimal value of $N$, and thus eliminating the stable components of
the error, is largest for the shortest assimilation windows: the largest
improvement, a $30\% $ reduction of the error with respect to classical 4DVar
is obtained for the smallest $\tau$ corresponding to one day, while the
improvement becomes less significant for larger $\tau$, and is about $20 \%$
for the five days window.

The experiments were repeated using larger values of $\sigma_o=1$ and
$\sqrt{2}$ and similar results were obtained (not shown) except that, as
expected, the error scales with $\sigma_o$.  It was not possible to complete
all the experiments with a window of five days and the larger value of
$\sigma_o$ because the algorithm failed due to increased nonlinearity; in
addition, the error minimum is shifted to slightly larger values of $N$:  these
were, for instance equal to $17$, $25$ and $30$ for the $40$, $60$ and $80$-variable
systems respectively when the largest observational error value
($\sigma_o=\sqrt{2}$) was used. This can be explained by noting that, when the
error of the initial guess is larger, also the estimate of the unstable
directions is less accurate at $t=t_0$; in such case a slightly larger
subspace is needed.

\subsubsection{Stable and unstable error components within the assimilation window}
\begin{figure*}
\includegraphics[scale=1.35]{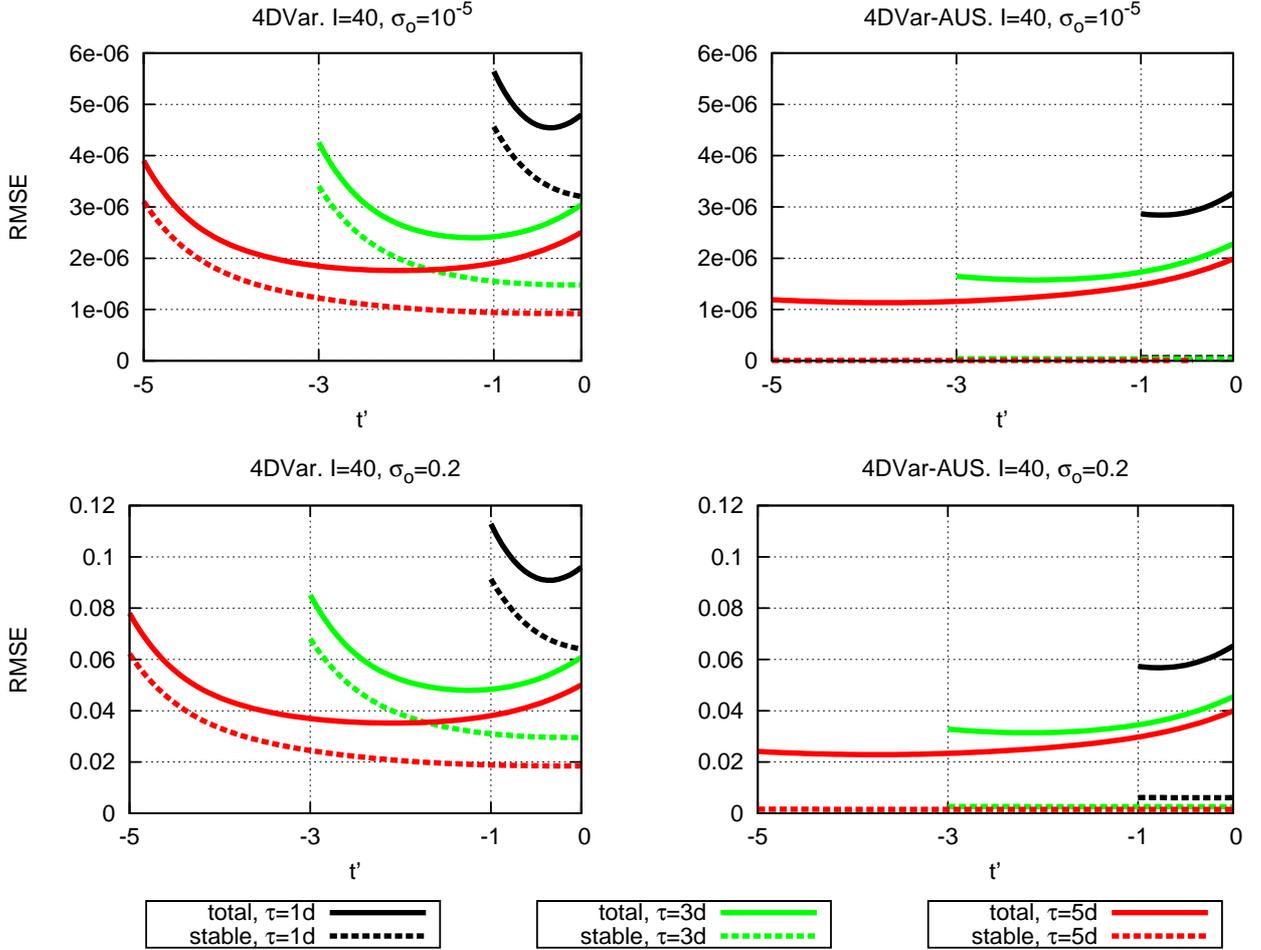} 
\caption{Time average RMS error within $1,3,5$ days assimilation windows
 as a function of $t'=t-\tau$, with  $\sigma_o=.2, 10^{-5}$
 for the model configuration $I=40$. Left panel: 4DVar. Right panel: 4DVar-AUS with $N=15$. Solid lines refer to total assimilation error, dashed lines refer to the error component in the stable subspace ${\bf e}_{16},...,{\bf e}_{40}$.}
\label{figura3}
\end{figure*}

The time dependence of the error within the assimilation window, shown in Fig. \ref{figura3}  
confirms the theoretical results of
Section \ref{teoria} and provides further insight on the reasons why 4DVar-AUS
performs better than 4DVar. In the figure we show the 4DVar and 4DVar-AUS
total error and the error projection on the stable directions ${\bf
e}_{N^++N^0+1},...,{\bf e}_I$, averaged over $5000$ consecutive assimilation windows.
The results shown were obtained with two values of $\sigma_o=10^{-5} $ and
$.2$, small enough that the analysis error scales with the observational error
as predicted by the tangent linear theory of Section \ref{teoria}.  

Figure \ref{figura3} shows that, according to the theory, the 4DVar error is
relatively larger in the stable subspace at initial time and in the unstable
and neutral subspace at final time.  
Instead, in 4DVar-AUS errors are very small  in the stable directions and
project almost totally on the unstable and neutral subspace.

The 4DVar-AUS assimilation error is smaller than the 4DVar error particularly 
for short assimilation windows.

Because the search of the minimum of the 4DVar cost function is conducted
in the entire phase space,  its minimum cannot be larger than the minimum of
the 4DVar-AUS cost function: this is confirmed by experimental evidence, the
4DVar cost function being typically a few percent smaller than the 4DVar-AUS
cost function.  We conclude that  the 4DVar solution is closer to the
observations but farther away from the real trajectory than the 4DVar-AUS
solution. This is due to the fact that, in 4DVar, errors in the stable
directions are ``kept alive'' by the observational error. In 4DVar-AUS, errors in
the stable directions being never corrected, are naturally damped along the
assimilation cycle: as a consequence, on  average errors project only  on the
long term Lyapunov vectors contained in the matrix ${\bf E}_0$. 

Results obtained by setting $\sigma_o=0$ show that the analysis error tends to
zero in a time span that is shorter for standard 4DVar than for 4DVar-AUS; thus
with perfect observations the full space 4DVar assimilation performs better, strengthening 
our conclusions.
It is worth mentioning that the evolution of ${\bf E}_0$ is a key factor
for the performance of the assimilation: in fact, experiments performed by using any
number $N<I$ of random directions show that the error is always larger
than the error of the full space 4DVar assimilation ($N=I$).

\subsubsection{Inclusion of the background term}
The 4DVar-AUS algorithm, in the absence of an explicit 
background term, amounts to assuming that the background error 
covariance matrix B is infinite in the unstable and neutral space, and $0$ 
in the stable subspace. The full 4DVar-AUS algorithm,  still 
in the absence of an explicit background term, amounts to 
assuming that the matrix B is globally infinite. 
For completeness, a set of experiments was conducted with the 
inclusion of an explicit finite background term in the 
cost function. The static background error covariance
matrix $\bf B$ was optimized for each assimilation window by the following
iterative procedure.  Starting from an initial guess, ${\bf B}$ is updated at
each iteration step with the covariance of the difference between the forecast
and true state, estimated from an assimilation cycle of 1000 consecutive
windows. The process is
repeated until convergence is obtained; in practice the iteration stops  when the analysis error,
averaged over the 1000 windows cycle, converges to an approximately constant value.
 To reduce the burden of computations 
$\bf B$ is optimized  for each window only for standard 4DVar ($N=I$);  the same matrix
is used for  the 4DVar-AUS experiments with the same window: in this way
4DVar-AUS is penalized, since its results could only improve if we used a matrix
$\bf B$ specifically optimized for each given subspace dimension $N<I$.

\begin{figure*}
\includegraphics[scale=0.435]{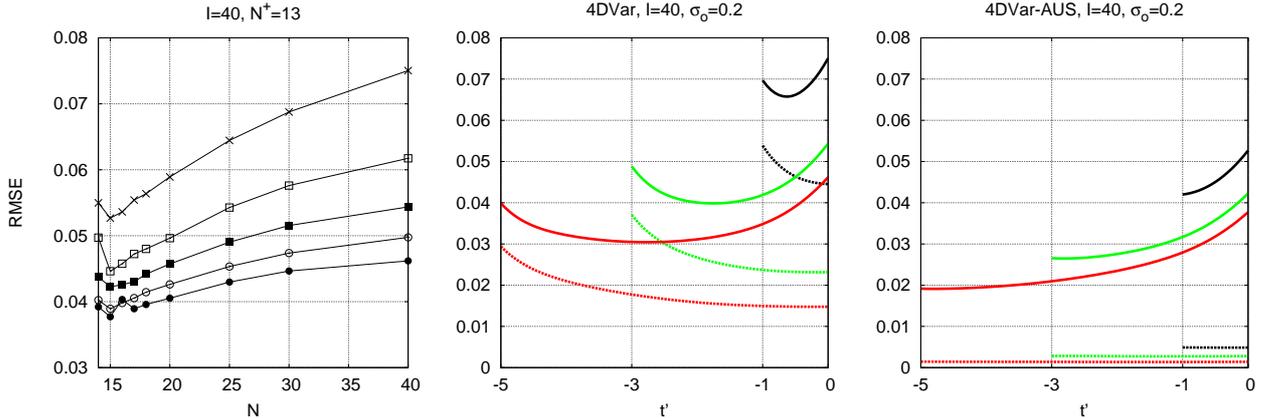}
\caption{Left, middle and right panels same as: left panel of Fig. \ref{fig1}
and bottom left and right panels of Fig. \ref{figura3} respectively, but for
experiments with background term in the cost function.  $\bf B$ was 
optimized as explained in the text.}
\label{fig4}
\end{figure*}

Results are shown in Fig. \ref{fig4}, to be compared with 
Fig. \ref{fig1} (for $I=40$) and with Fig. \ref{figura3} 
(for $\sigma_o=0.2$). Everything else being equal, the 
introduction of the background term leads to an overall 
improvement of the performance in all experiments (compare 
the middle panel of Fig. \ref{fig4} with the the lower left 
panel of Fig. \ref{figura3} as concerns full 4DVar, and the 
right panel of Fig. \ref{fig4} with the lower right panel 
of Fig. \ref{figura3} as concerns 4DVar-AUS). This shows 
that useful information is contained in the matrix $\bf B$. 
The most accurate analysis is still obtained for a value of 
$N$ that is just above the number of positive and null Lyapunov 
exponents. It is seen that, in agreement with the theory, the largest
improvements are obtained for shorter windows and for standard 4DVar. As
expected, increasing the length of the time window decreases the influence of
the background term. For 4DVar, the effect of the background term is
particularly efficient in reducing the analysis error along the stable
directions. Therefore a significant reduction of the error at the beginning of
the time window is observed, but errors are nevertheless still present in the
analysis in the stable directions.  The conclusions to be drawn as to the
matrix $\bf B$ as it has been defined here are first, as said, that it contains
useful information in both the stable and unstable subspaces.  Now, the error
in the stable subspace is only partially decreased in the full 4DVar, while it
is entirely eliminated in 4DVar-AUS. So, the description that the matrix $\bf
B$ gives of the error in the stable subspace is more accurate than assuming
that this error is infinite, but less accurate than assuming it is zero.

\section{Conclusions}
\label{conclusions}
One of the main purposes of the present paper has been the development of
four-dimensional variational assimilation in the unstable subspace.  The
results provide a proof-of-concept, at least in the case of the simple model
used in this study, of the benefit in terms of assimilation performance of
selecting the subspace where instabilities develop.  The key result of this
study is the existence of an optimal subspace dimension for the assimilation
that is directly related to the unstable and neutral subspace dimension.  The
selected subspace - the leading Lyapunov vectors subspace -  contains the most
rapidly growing perturbations.  In the presence of observational error, the
optimal number of directions is approximately equal to $N^++N^0$, where $N^+$
and $N^0$ are the number of positive and null Lyapunov exponents. The 4DVar
solution ($N=I$), while being closer to the observations, is farther away from
the truth.  This result has been explained showing that, when we assimilate in
the unstable and neutral subspace, errors in the stable directions are
naturally damped.  Because of observational error, assimilating in the whole
space otherwise keeps the stable components of the error alive, deteriorating
the overall assimilation performance. If the observational error is zero, the
optimal dimension is the dimension $I$ of the whole space.

The present theoretical results can have implications for the application of
advanced assimilation methods to high-dimensional models of the atmosphere and
ocean. Here, we have shown that 4DVar could benefit from
the dynamical information on the unstable directions - the ``optimal'' subspace
where the analysis error is confined - propagated along the assimilation cycle.
The possible application of the present findings to more realistic contexts is
left for future investigations.  Work is in progress to explore the existence
of an optimal subspace dimension for EnKFs.

The results presented in this paper have been obtained with a simple and
economical numerical model and, strictly speaking, do not prove anything as to
what would be obtained with a more realistic model of the atmospheric flow. At
the same time, the considerations which have led to the design of the
experiments described in the paper, and which are confirmed by the results that
have been obtained, are very general, and do not fundamentally require anything
else than the existence of both stable and unstable modes in the system?? under
consideration. One possible source of difficulties could be the unavoidable
presence of errors in the assimilating model.  The experiments described here
have been performed under the hypothesis of a perfect model. In the more
realistic situation of an imperfect model, the corresponding errors will modify
the unstable subspace, at least to some extent. The results of
\citet{trevisan_etal-jas-2004} showed that the performance of AUS assimilation
is not severely affected by the presence of model error in the model used
above. Further work will be necessary in order to assess to which degree the
presence of model errors in more realistic assimilation problems can affect the
conclusions that have been obtained here. 

%\subsection{Acknowledgements} An Acknowledgements section is started with \verb"\ack" or
%\verb"\acks" for \textit{Acknowledgement} or
%\textit{Acknowledgements}, respectively. It must be placed just
%before the References (or before the appendix when applicable).
%\subsection{Acknowledgements} An Acknowledgements section is started with \verb"\ack" or

%\acks
%The authors thank their colleagues Alberto Maurizi and Maurizio Fantini for
%useful discussions and Alberto Carrassi and Francesco Uboldi for their
%continuing interest in the development of AUS.

\appendix
\section{Effect of the confinement on the efficiency of the numerical algorithm}
\label{appendice}

The minimization in the reduced subspace, chosen to be
spanned by the most unstable directions is expected to converge more rapidly
in view of the following argument. The Hessian, under the same simplifying
hypothesis used to derive the analysis error covariance (\ref{c0_pires}) can 
be written as:

\begin{equation}
\label{Hess}
\begin{split}
  \widetilde{\frac{d^2 J}{d{\bf x}_0^2}}= \sigma_o^{-2}{\bf E}_0 diag \biggl[&\sum_{i=0}^{n}( {\bf \Lambda}_i^{(1)})^2, \sum_{i=0}^{n}( {\bf \Lambda}_i^{(2)})^2,...,\\
&,...,\sum_{i=0}^{n}( {\bf \Lambda}_i^{(N)})^2\biggr]{\bf E}_0^T
\end{split}
\end{equation}

The condition number, ratio of the largest to the smallest eigenvalue of (\ref{Hess}), can be  reduced if $N$ is significantly smaller than the model space dimension.

\bibliographystyle{isac-abl}
\bibliography{bibliografia.bib}

\begin{thebibliography}{37}
\expandafter\ifx\csname natexlab\endcsname\relax\def\natexlab#1{#1}\fi

\bibitem[Bannister, 2008]{bannister-qjrms-2008}
Bannister, R.~N., 2008: A review of forecast error covariance statistics in
  atmospheric variational data assimilation. {I}: Characteristics and
  measurements of forecast error covariances. \textit{Q. J. Roy. Meteor. Soc.},
  \textbf{134}, 1951--1970.

\bibitem[Benettin et~al., 1980]{benettin_etal-meccanica-1980}
Benettin, G., L.~Galgani, A.~Giorgilli, and J.~M. Strelcyn, 1980: Lyapunov
  characteristic exponents for smooth dynamical systems and for {H}amiltonian
  systems; a method for computing them. \textit{Meccanica}, \textbf{15}, 9--21.

\bibitem[Brown et~al., 1991]{brown_etal-pra-1991}
Brown, R., P.~Bryant, and H.~D.~I. Abarbanel, 1991: Computing the {L}yapunov
  spectrum of a dynamical system from an observed time series. \textit{Phys.
  Rev. A}, \textbf{43}.

\bibitem[Carrassi et~al., 2008{\natexlab{a}}]{carrassi_etal-chaos-2008}
Carrassi, A., M.~Ghil, A.~Trevisan, and F.~Uboldi, 2008{\natexlab{a}}: Data
  assimilation as a nonlinear dynamical systems problem: {S}tability and
  convergence of the prediction-assimilation system. \textit{Chaos},
  \textbf{18}.

\bibitem[Carrassi et~al., 2008{\natexlab{b}}]{carrassi_etal-npg-2008}
Carrassi, A., A.~Trevisan, L.~Dechamps, O.~Talagrand, and F.~Uboldi,
  2008{\natexlab{b}}: Controlling instabilities along a 3{DV}ar analysis cycle
  by assimilating in the unstable subspace: a comparison with the {E}n{KF}.
  \textit{Nonlinear Proc. Geoph.}, \textbf{15}, 503--521.

\bibitem[Carrassi et~al., 2007]{carrassi_etal-tellus-2007}
Carrassi, A., A.~Trevisan, and F.~Uboldi, 2007: Adaptive observations and
  assimilation in the unstable subspace by breeding on the data-assimilation
  system. \textit{Tellus}, \textbf{59}, 101--113.

\bibitem[Daley, 1991]{daley-1991}
Daley, R., 1991: \textit{Atmospheric Data Analysis}, Cambridge University
  Press.

\bibitem[Eckmann and Ruelle, 1985]{eckmann_etal-rmp-1985}
Eckmann, J.~P. and D.~Ruelle, 1985: Ergodic-theory of chaos and strange
  attractors. \textit{Rev. Mod. Phys.}, \textbf{57}, 617--656.

\bibitem[Evensen, 1994]{evensen-jgr-1994}
Evensen, G., 1994: Sequential data assimilation with a nonlinear
  quasi-geostrophic model using monte carlo methods to forecast error
  statistics. \textit{J. Geophys. Res.}, \textbf{99}, 10143--10162.

\bibitem[Fertig et~al., 2007]{fertig_etal-tellus-2007}
Fertig, E.~J., J.~Harlim, and B.~R. Hunt, 2007: A comparative study of
  4{D}-{VAR} and a 4{D} {E}nsemble {K}alman {F}ilter: perfect model simulations
  with {L}orenz-96. \textit{Tellus}, \textbf{95A}, 96--100.

\bibitem[Ghil and Malanotte~Rizzoli, 1991]{ghil_etal-ag-1991}
Ghil, M. and P.~Malanotte~Rizzoli, 1991: Data assimilation in meteorology and
  oceanography, Academic Press, pp. 141--266.

\bibitem[Gustafsson, 2007]{gustafsson-tellus-2007}
Gustafsson, N., 2007: Discussion on `4{D}-{V}ar or {E}n{KF}?'. \textit{Tellus},
  \textbf{59A}, 774--777.

\bibitem[Hamill et~al., 2001]{hamill_etal-mwr-2001}
Hamill, T.~M., J.~S. Whitaker, and C.~Snyder, 2001: {D}istance-{D}ependent
  {F}iltering of {B}ackground {E}rror {C}ovariance {E}stimates in an {E}nsemble
  {K}alman {F}ilter. \textit{Mon. Weather Rev.}, \textbf{129}, 2776--2790.

\bibitem[Jazwinski, 1970]{jazwinski-1970}
Jazwinski, A.~H., 1970: \textit{Stochastic {P}rocesses and {F}iltering
  {T}heory}, Academic Press.

\bibitem[Kalnay, 2003]{kalnay-2003}
Kalnay, E., 2003: \textit{Atmospheric {M}odeling, {D}ata {A}ssimilation and
  {P}redictability}, Cambridge University Press.

\bibitem[Kalnay et~al., 2007]{kalnay_etal-tellus-2007}
Kalnay, E., H.~Li, T.~Miyoshi, S.-C. Yang, and J.~Ballabrera-Poy, 2007:
  4-{D}-{V}ar or ensemble {K}alman filter? \textit{Tellus}, \textbf{59A},
  758--773.

\bibitem[Le~Dimet and Talagrand, 1986]{ledimet_etal-tellus-1986}
Le~Dimet, F.~X. and O.~Talagrand, 1986: Variational algorithms for analysis and
  assimilation of meteorological observations - theoretical aspects.
  \textit{Tellus}, \textbf{38}, 97--110.

\bibitem[Legras and Vautard, 1996]{legras_etal-1996}
Legras, B. and R.~Vautard, 1996: A guide to {L}yapunov vectors, \textit{Proc.
  Seminar on Predictability, Vol. 1}, ECMWF, Reading, Berkshire, UK.

\bibitem[Lorenc, 2003]{lorenc-qjrms-2003}
Lorenc, A.~C., 2003: The potential of the e {E}nsemble {K}alman {F}ilter for
  {N}{W}{P} - a comparison with {4D-VAR}. \textit{Q. J. Roy. Meteor. Soc.},
  \textbf{129}, 3183--3293.

\bibitem[Lorenz, 1984]{lorenz-physica-1984}
Lorenz, E.~N., 1984: The local structure of a chaotic attractor in
  four-dimensions. \textit{Physica D}, \textbf{13}, 90--104.

\bibitem[Lorenz, 1996]{lorenz-1996}
Lorenz, E.~N., 1996: Predictability: {A} problem partly solved, \textit{Proc.
  Seminar on Predictability, Vol. 1}, ECMWF, Reading, Berkshire, UK.

\bibitem[Lorenz and Emanuel, 1998]{lorenz_etal-jas-1998}
Lorenz, E.~N. and K.~A. Emanuel, 1998: Optimal {S}ites for {S}upplementary
  {W}eather {O}bservations: {S}imulation with a {S}mall {M}odel. \textit{J.
  Atmos. Sci.}, \textbf{55}, 399--414.

\bibitem[Miller et~al., 1994]{miller_etal-jas-1994}
Miller, R.~N., M.~Ghil, and F.~Gauthiez, 1994: Advanced data assimilation in
  strongly nonlinear dynamical systems. \textit{J. Atmos. Sci.}, \textbf{51},
  1037--1056.

\bibitem[Oseledec, 1968]{oseledec-1968}
Oseledec, V.~I., 1968: A {M}ultiplicative {E}rgodic {T}heorem, {L}yapunov
  {C}haracteristic {N}umbers for {D}ynamical {S}ystems. \textit{Trans. Moscow
  Math. Soc.}, \textbf{19}, 197--231.

\bibitem[Pires et~al., 1996]{pires_etal-tellus-1996}
Pires, C., R.~Vautard, and O.~Talagrand, 1996: On extending the limits of
  variational assimilation in nonlinear chaotic systems. \textit{Tellus},
  \textbf{48A}, 96--121.

\bibitem[Press et~al., 1992]{num_recipes}
Press, W.~H., S.~A. Teukolsky, W.~T. Vetterling, and B.~P. Flannery, 1992:
  \textit{Numerical Recipes in {FORTRAN}}, 2nd ed., Cambridge University Press.

\bibitem[Swanson et~al., 2000]{swanson_etal-jas-2000}
Swanson, K.~L., T.~N. Palmer, and R.~Vautard, 2000: Observational {E}rror
  {S}tructures and the {V}alue of {A}dvanced {A}ssimilation {T}echniques.
  \textit{J. Atmos. Sci.}, \textbf{57}, 1327--1340.

\bibitem[Talagrand, 1997]{talagrand-jmsj-1997}
Talagrand, O., 1997: Assimilation of observations, an introduction. \textit{J.
  Meteorol. Soc. Jpn.}, \textbf{75}, 191--209.

\bibitem[Talagrand and Courtier, 1987]{talagrand_etal_qjrms-1987}
Talagrand, O. and P.~Courtier, 1987: Variational assimilation of observations
  with the adjoint vorticity equations. \textit{Q. J. Roy. Meteor. Soc.},
  \textbf{113}, 1311--1328.

\bibitem[Toth and Kalnay, 1993]{toth_etal-bams-1993}
Toth, Z. and E.~Kalnay, 1993: Ensemble forecasting at {N}{M}{C}: {T}he
  {G}eneration of {P}erturbations. \textit{B. Am. Meteorol. Soc.}, \textbf{74},
  2317--2330.

\bibitem[Toth and Kalnay, 1997]{toth_etal-mwr-1997}
Toth, Z. and E.~Kalnay, 1997: Ensemble forecasting at {NCEP} and the breeding
  method. \textit{Mon. Weather Rev.}, \textbf{125}, 3297--3319.

\bibitem[Trevisan and Pancotti, 1998]{trevisan_etal-jas-1998}
Trevisan, A. and F.~Pancotti, 1998: Periodic {O}rbits, {L}yapunov {V}ectors,
  and {S}ingular {V}ectors in the {L}orenz {S}ystem. \textit{J. Atmos. Sci.},
  \textbf{55}, 390--398.

\bibitem[Trevisan and Uboldi, 2004]{trevisan_etal-jas-2004}
Trevisan, A. and F.~Uboldi, 2004: Assimilation of {S}tandard and {T}argeted
  {O}bservations within the {U}nstable {S}ubspace of the
  {O}bservation-­{A}nalysis-­{F}orecast {C}ycle {S}ystem. \textit{J. Atmos.
  Sci.}, \textbf{61}, 103--113.

\bibitem[Tsuyuki and Miyoshi, 2007]{tsuyuki_etal-jmsj-2007}
Tsuyuki, T. and T.~Miyoshi, 2007: Recent {P}rogress of {D}ata {A}ssimilation
  {M}ethods in {M}eteorology. \textit{J. Meteorol. Soc. Jpn.}, \textbf{85B},
  331--361.

\bibitem[Uboldi and Trevisan, 2006]{uboldi_etal-npg-2006}
Uboldi, F. and A.~Trevisan, 2006: Detecting unstable structures and controlling
  error growth by assimilation of standard and adaptive observations in a
  primitive equation ocean model. \textit{Nonlinear Proc. Geoph.}, \textbf{13},
  67--81.

\bibitem[Whitaker and Hamill, 2002]{whitaker_etal-mwr-2002}
Whitaker, J.~S. and T.~M. Hamill, 2002: Ensemble data assimilation without
  perturbed observations. \textit{Mon. Weather Rev.}, \textbf{130}, 1913--1924.

\bibitem[Wolfe and Samelson, 2007]{wolfe_etal-tellus-2007}
Wolfe, C.~L. and R.~M. Samelson, 2007: An efficient method for recovering
  {L}yapunov vectors from singular vectors. \textit{Tellus A}, \textbf{59},
  355--366.

\end{thebibliography}

\end{document}